\documentclass[twoside,a4paper,11pt]{sca}
\usepackage{graphicx}
\usepackage{hyperref}
\usepackage{movie15}
\begin{document}
\pagenumbering{arabic}
\pagestyle{myheadings}
\thispagestyle{empty}
{\flushright\includegraphics[width=\textwidth,bb=90 650 520 700]{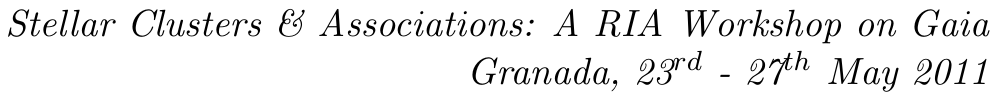}}
\vspace*{0.2cm}
\begin{flushleft}
{\bf {\LARGE
%
The field of  Lod\'{e}n 112
%
}\\
\vspace*{1cm}
%
N. Kaltcheva$^{1}$
and
V. Golev$^{2}$
%
}\\
\vspace*{0.5cm}
%
$^{1}$
Department of Physics and Astronomy, University of Wisconsin Oshkosh,
  800 Algoma Blvd., Oshkosh, WI 54901, USA (kaltchev@uwosh.edu)\\
$^{2}$
Department of Astronomy, Faculty of Physics, St Kliment Ohridski University of
Sofia, 5 James Bourchier Blvd., BG-1164 Sofia, Bulgaria (valgol@phys.uni-sofia.bg)\\
%
\end{flushleft}
%
\markboth{
Lod\'{e}n 112
}{ 
%
Kaltcheva \& Golev  
%
}
\thispagestyle{empty}
\vspace*{0.4cm}
\begin{minipage}[l]{0.09\textwidth}
\ 
\end{minipage}
\begin{minipage}[r]{0.9\textwidth}
\vspace{1cm}
\section*{Abstract}{\small
%
Based on the available $uvby\beta$
  photometry of OB stars in the longitude range $281^\circ$ to $285^\circ$ in
  the Galactic disk, we identify a feature of young stars at 1630$\pm82$ pc,
  that is probably connected to the compact cluster candidate
  Lod\'{e}n 112 and the open cluster IC 2581. This feature seems
  to be spatially correlated to RCW 48 and RCW 49 and several other smaller
  H{\sc ii} regions.
%
\normalsize}
\end{minipage}
%
%
%
\section{Introduction \label{intro}}
The field between the OB groups in Vela ($262^\circ<l<268^\circ$) and
Car~OB1 ($284^\circ<l<288^\circ$) is not known to be dominated by any
prominent OB association (see \cite{humphreys78}, \cite{melnik95}). The longitude
range $283^\circ$\,-\,$284^\circ$ is thought to correspond to a tangential
direction of a large segment of the Carina arm in both the 3- and 4-arm models
of the grand design of the Milky Way \cite{russeil03}. Our study is based on
$uvby\beta$ photometric distances and provides new insights on the apparent
groupings and layers toward $281^\circ<l<285^\circ$ in the Galactic
disk. This field clearly stands apart from the extended H{\sc ii} features
toward Car OB1 and contains several prominent H{\sc ii} regions.

\section{Results}
\begin{figure}
\center
\includegraphics[scale=0.31]{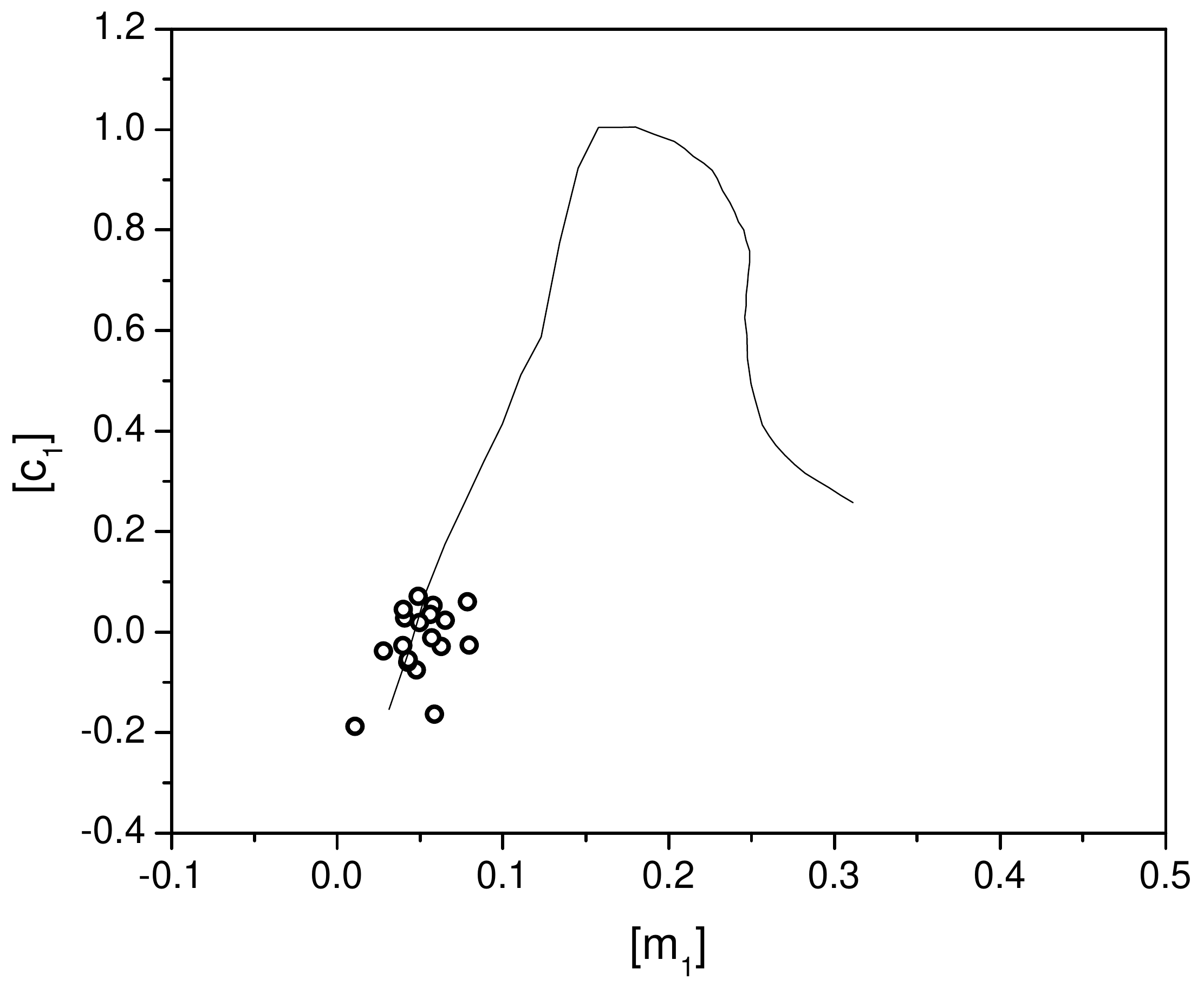} ~
\includegraphics[scale=0.5]{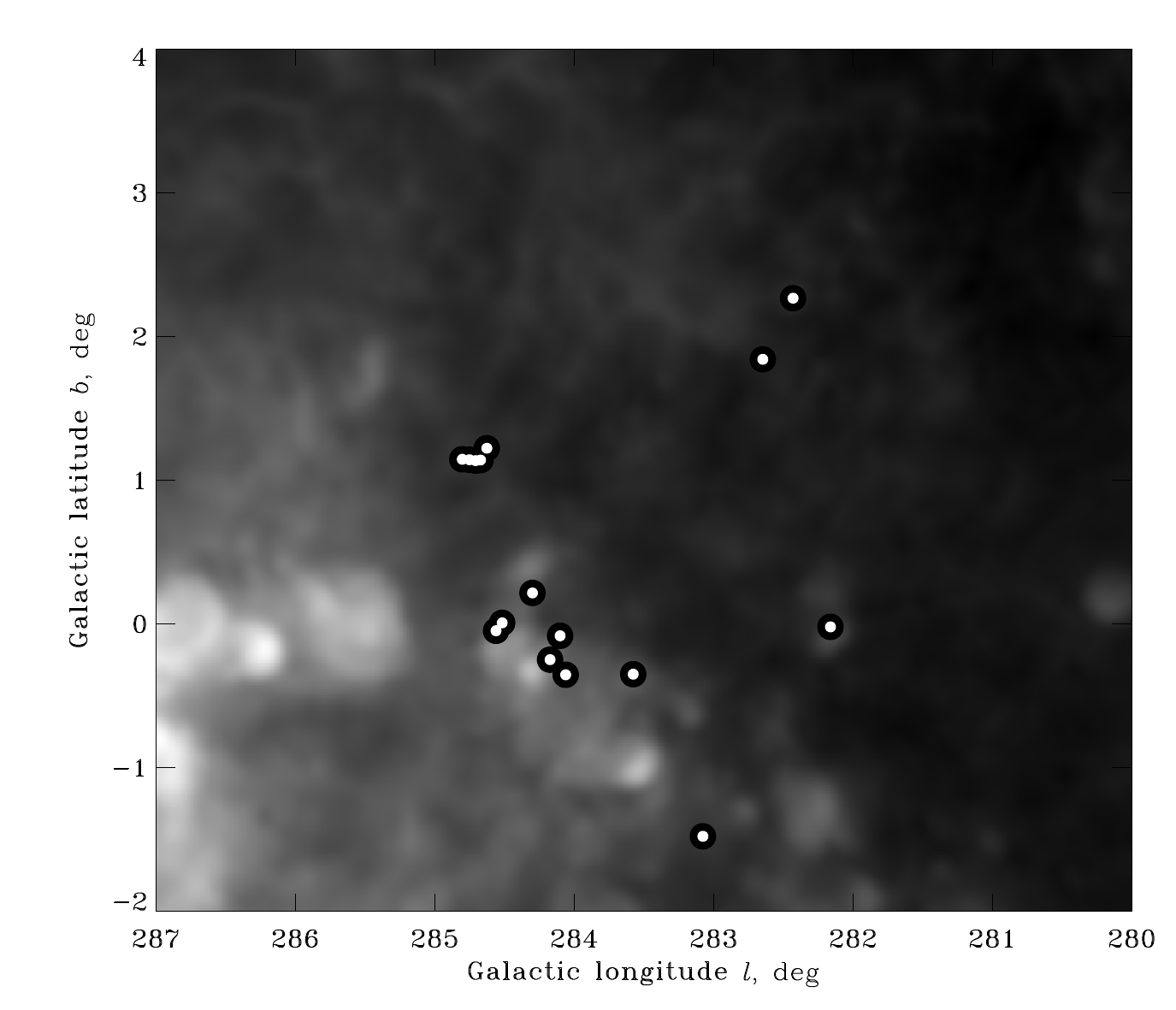} 
\caption{\label{fig1} Left: the classification photometric diagram for the OB
  stars in the sample. Right:  same stars overplotted on the distribution of the
  H{\sc ii} emission in the region \cite{finkbeiner03} obtained via {\em SkyView} 
  VO interface \cite{McGlynn98}.}
\end{figure}

 Our sample includes a number of OB stars with very similar luminosities
 according to the $[c_{1}]$ vs. $[m_{1}]$  classification diagram
 \cite{stromgren66} (Fig.~\ref{fig1}, left). Among them 5 stars belong to the
 poor but very compact cluster candidate Lod\'{e}n 112. Two
 other stars are assigned to the relatively young cluster IC 2581, and the
 rest are apparent field stars. According to the $uvby\beta$ photometric
 distances all OB stars in this sample show a very small spread in distance,
 thus forming a structure at 1630$\pm$80 pc.  This new estimate is
 significantly smaller than the presently adopted distance of 2500 pc to
 Lod\'{e}n 112 (WEBDA database). These OB-stars are located
 toward the prominent H{\sc ii} regions RCW 48 and RCW 49 and several smaller
 H{\sc ii} fields and could represent a new OB association at coordinates
 $282^\circ\!<l<285^\circ\!$, $-2^\circ\!<b<2^\circ\!$ (Fig.~\ref{fig1},
 right), containing the compact cluster candidate Lod\'{e}n 112 
 and IC 2581 (more details can be found in \cite{kaltcheva11}).

%
%
\small  
%
\section*{Acknowledgments}   
%
This work is supported by the National Science Foundation grant
AST-0708950.  N.K. acknowledges support from the SNC Endowed Professorship
at the University of Wisconsin Oshkosh.  V.G. acknowledges support by
the Bulgarian National Science Research Fund grants DO 02-85/2008 and
DO 02-362/2008. This research has made use of the SIMBAD database, operated
at CDS, Strasbourg, France. We acknowledge the use of NASA's {\em SkyView}
facility (http://skyview.gsfc.nasa.gov).


%
\end{document}